\documentclass[11pt, twoside]{article}

\usepackage{asp2010}

\begin{document}
\title{Linking to Data - Effect on Citation Rates in Astronomy}   %%% Fill in title
\author{Edwin A. Henneken$^1$, Alberto Accomazzi$^1$}   %%% Fill in author names
\affil{$^1$Smithsonian Astrophysical Observatory, 60 Garden Street, Cambridge, MA 02138}    %%% Fill in author affiliations

\begin{abstract} %%% Abstract to run on from here.

Is there a difference in citation rates between articles that were published with links to data and articles that were not? Besides being interesting from a purely academic point of view, this question is also highly relevant for the process of furthering science. Data sharing not only helps the process of verification of claims, but also the discovery of new findings in archival data. However, linking to data still is a far cry away from being a ``practice'', especially where it comes to authors providing these links during the writing and submission process. You need to have both a willingness and a publication mechanism in order to create such a practice. Showing that articles with links to data get higher citation rates might increase the willingness of scientists to take the extra steps of linking data sources to their publications. In this presentation we will show this is indeed the case: articles with links to data result in higher citation rates than articles without such links. 

The ADS is funded by NASA Grant NNX09AB39G.
\end{abstract}

\section{Introduction}
Furthering science depends to a large degree on knowledge and information transfer. Therefore it critically relies on discoverability. This applies to findings in publications and to the underlying data that led to these findings. Therefore, significant amounts of energy (and funds) should be invested in improving discoverability, of both publications and data. Major progress has been made on the level of publications by improved visibility and more sophisticated techniques for information discovery. The adoption of faceted filtering, recommender systems and semantic interlinking of resources are good examples of this (\citet{accomazzi11}, \citet{henneken11}). 

It is time that exposure of data becomes common practice. A publication based on a data set is just one expression of the potential of that data set. It totally depends on the background and the interests of the researchers which representation of that potential will be selected. However, there are many other representations. The scientific community would also benefit greatly from the ability to combine a data set with other available data sets. Also, having data available publicly would greatly facilitate the verification of claims (\citet{fischer10}). The special session ``The Literature-Data Connection: Meaning, Infrastructure and Impact'' at the 218th Meeting of the American Astronomical Society (Boston, May 2011) was dedicated to this discussion. As part of the discussion of how to create a practice of linking data to publications, the question was raised whether such publications would see a citation advantage. That would be like getting a tax benefit for ``being green''. Everybody agrees that ``being green'' is a sensible thing to do, but having some kind of incentive definitely helps as additional motivation. Motivation is an essential ingredient for creating a practice. Since citations are a measure used for scientific impact, it is logical to ask whether investing energy into making data available publicly results in a citation advantage.

In this presentation we address the question whether there is a citation advantage. We explore the question using the holdings and citation data of the SAO/NASA Astrophysics Data System (ADS).
\section{Results}
\begin{figure}[!ht]
  \plotone{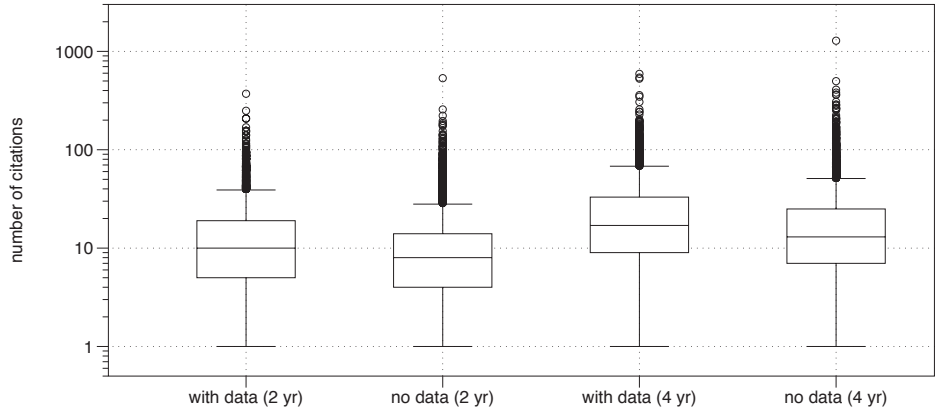}
  \caption{Distribution of citations of articles published in \emph{The Astrophysical Journal} (including \emph{Letters} and \emph{Supplement}), \emph{The Astronomical Journal}, \emph{The Monthly Notices of the R.A.S.} and \emph{Astronomy \& Astrophysics} including \emph{Supplement}), during the period 1995 through 2000. The extent of the box corresponds with the interquartile range of the citations and whiskers extend to 1.5 times the interquartile range. The horizontal lines within the boxes correspond with the medians. From left to right, the boxes correspond respectively with the citation distributions for the article set with and without data links 2 years after publication, and 4 years after publication. The medians are respectively at 10, 8, 17 and 13 citations.}
  \label{fig:BoxDiagram}
\end{figure}

\begin{figure}[!ht]
  \plotone{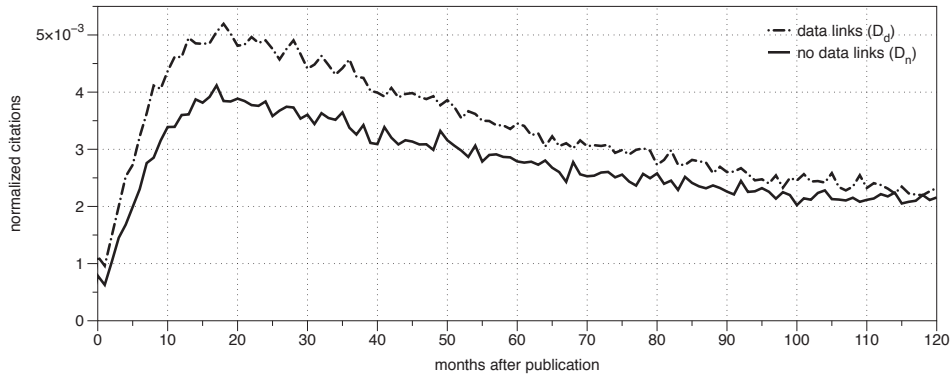}
  \caption{The normalized number of citations for data sets $\emph{D}_d$ and $\emph{D}_n$. The citations have been normalized by the total number of citations.}
  \label{fig:normalizedCites}
\end{figure}

\begin{figure}[!ht]
  \plotone{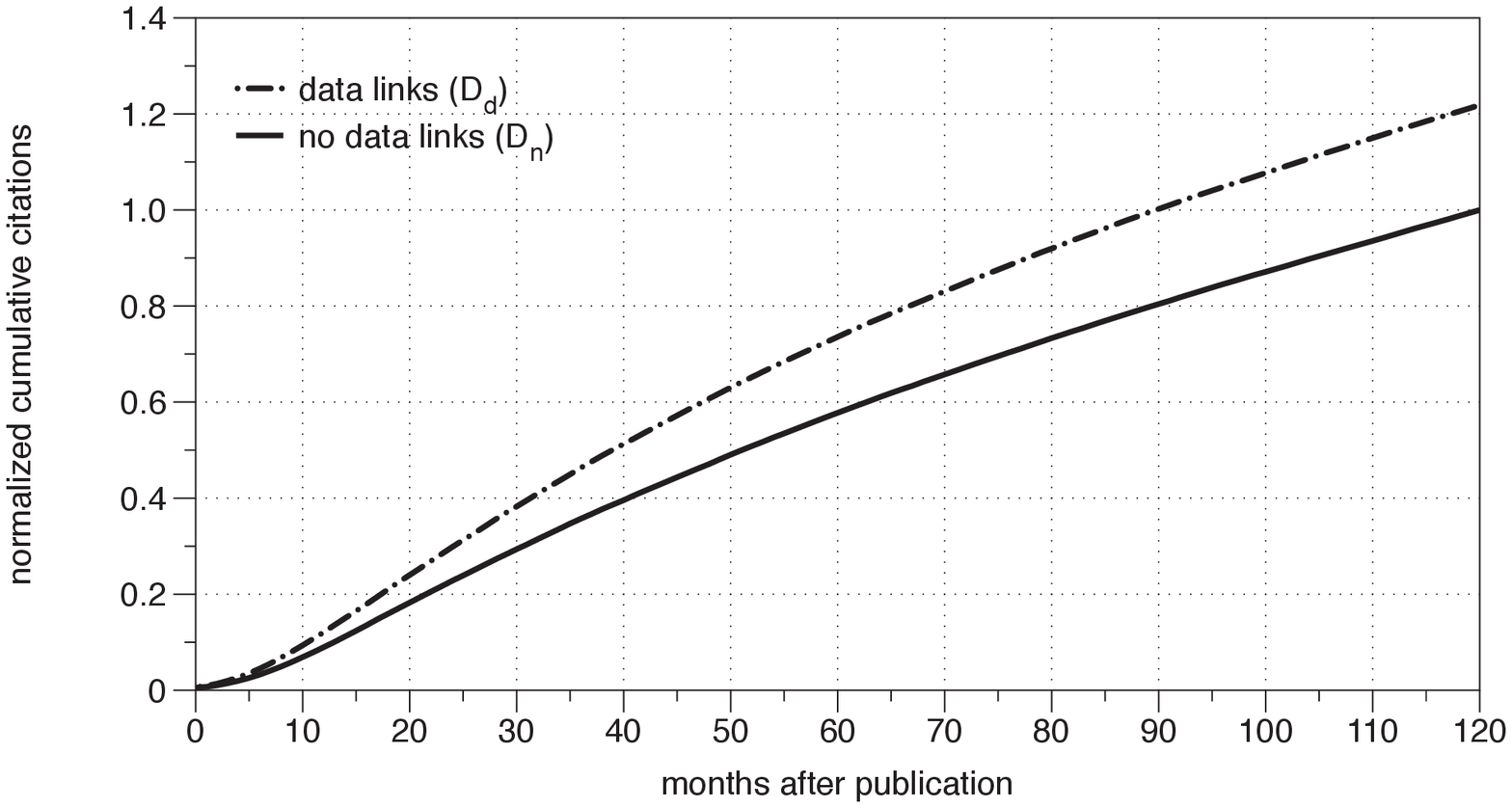}
  \caption{The cumulative citation distributions for data sets $\emph{D}_d$ and $\emph{D}_n$. The citation counts have been normalized by the total number of citations for articles without data links, 10 years after publication.}
  \label{fig:cumulativeCites}
\end{figure}

With every record in the ADS holdings a number of possible attributes (``links'') can be associated, giving access to information related to that record. The attribute used for this analysis is the ``D'' link, associated with access to on-line data. Currently these links point to data hosted at data centers (like CDS, HEASARC and MAST). The following set of records was chosen for this study: articles published in \emph{The Astrophysical Journal} (including \emph{Letters} and \emph{Supplement}), \emph{The Astronomical Journal}, \emph{The Monthly Notices of the R.A.S.} and \emph{Astronomy \& Astrophysics} including \emph{Supplement}), during the period 1995 through 2000. Comparing publications with a ``D'' link to those without such a link would, to a large degree, be comparing apples with oranges, because of the range in subject matter. In order for the comparison to make sense, the subject matter of the publications needs to be restricted. We decided to use keywords as filter. We determined the set of 50 most frequently used keywords in articles with data links. The articles to be used for the analysis were obtained by requiring that they have at least 3 keywords in common with that set of 50 keywords. This resulted in a set of 3814 articles with data links and 7218 articles without data links. The box diagram in figure~\ref{fig:BoxDiagram} characterizes the distribution of citations in the sets with and without data links, for respectively 2 and 4 years after publication. 
For this analysis, a random selection of 3814 articles was extracted from the set of 7218 articles (without links to data). For both sets the citation accumulation was determined for each article. From now on, we will refer to the set with data links as $\emph{D}_d$ and the one without data links as $\emph{D}_n$. These citation distributions were used to calculate the mean citation accumulation for each set, normalized by the total number of citations in the entire set of publications. The results are shown in figure~\ref{fig:normalizedCites}.
Figure~\ref{fig:normalizedCites} indicates that publications with a data link have a larger citation rate than publications that do not. To get get an indication of how much more citations a publication with a data link accumulates, on average, figure~\ref{fig:cumulativeCites} shows the cumulative citation distribution, normalized by the total number of citations for articles without data links, 10 years after publication.
Figure~\ref{fig:cumulativeCites} indicates that for this data set, articles with data links on average acquired 20\% more citations (compared to articles without these links) over a period of 10 years. The fact that this increase is statistically significant follows from a regression analysis performed on the entire data set. This confirmed the increase of 20\% in citation count (at a 95\% confidence level).
\section{Discussion}
Our study seems to indicate that publications with links to on-line data seem to have a higher citation rate than publications that do not. Could this effect be attributed to another systematic effect? For example, studies have shown that e-printing results in higher citation rates (see for example~\citet{henneken06}). However, both sets used to construct figures~\ref{fig:normalizedCites} and~\ref{fig:cumulativeCites} turn out to be homogeneous in other publication attributes. For example, in each set about 20\% of the publications have e-prints associated with them. So, the increased citation rates associated with e-printing contribute similarly in both sets. Also, both sets are homogenous in links to object information (NED and SIMBAD links). Lastly, could data centers, in attributing data links to articles, have cherry-picked important (i.e. more citable) data sets? Both sets of publications turn out to be homogenous in citation distributions as well. This leads us to believe that the effect observed is real.

In a study of medical literature on cancer microarray clinical trials, \citet{piwowar07} found that ``publicly available data was significantly associated with a 69\% increase in citations''. Even though citation rates are different for different disciplines, the qualitative observation still holds. Studies and discussions in other disciplines show that data sharing is viewed as important and highly relevant for the integrity and furthering of science, and that the hurdles encountered have much in common between various disciplines (\citet{bruna10}, \citet{delamothe96}, \citet{kansa10}, \citet{pisani10}, \citet{south10}, \citet{vickers11}, \citet{vandewalle09}). 
%%%\acknowledgements

\bibliographystyle{asp2010}
\bibliography{O20}{}

%\begin{thebibliography}{}

%\end{thebibliography}

\end{document}